\documentclass[conference]{IEEEtran}
\IEEEoverridecommandlockouts
\usepackage{cite}
\usepackage{amsmath,amssymb,amsfonts}
\usepackage{algorithmic}
\usepackage{graphicx}
\usepackage{textcomp}
\usepackage{xcolor}
\usepackage{algorithm}
\usepackage{booktabs}
\usepackage{graphicx}
\usepackage{subcaption}
\usepackage{hyperref}
\hypersetup{
    colorlinks=true,
    linkcolor=blue,
    filecolor=blue,      
    urlcolor=blue,
    citecolor=blue,
}
\usepackage{siunitx}

\DeclareRobustCommand*{\IEEEauthorrefmark}[1]{%
    \raisebox{0pt}[0pt][0pt]{\textsuperscript{\footnotesize\ensuremath{#1}}}}

\begin{document}

\title{Low-Latency Layer-Aware Proactive and Passive Container Migration in Meta Computing
}

\author{
\IEEEauthorblockN{
Mengjie Liu\IEEEauthorrefmark{2,1},
Yihua Li\IEEEauthorrefmark{5},
Fangyi Mou\IEEEauthorrefmark{3},
Zhiqing Tang\IEEEauthorrefmark{1},
Jiong Lou\IEEEauthorrefmark{4},
Jianxiong Guo\IEEEauthorrefmark{1,3},
and Weijia Jia\IEEEauthorrefmark{1,3}}
\IEEEauthorblockA{\IEEEauthorrefmark{1}Institute of Artificial Intelligence and Future Networks, Beijing Normal University, China}
\IEEEauthorblockA{\IEEEauthorrefmark{2}Faculty of Arts and Sciences, Beijing Normal University, China}
\IEEEauthorblockA{\IEEEauthorrefmark{3}Guangdong Key Lab of AI and Multi-Modal Data Processing, BNU-HKBU United International College, China}
\IEEEauthorblockA{\IEEEauthorrefmark{4}Department of Computer Science and Engineering, Shanghai Jiao Tong University, China}
\IEEEauthorblockA{\IEEEauthorrefmark{5}Department of Informatics, University of Zurich, Zurich, Switzerland}
\IEEEauthorblockA{mengjieliu@mail.bnu.edu.cn, yihua.li@uzh.ch, moufangyi@uic.edu.cn, \\zhiqingtang@bnu.edu.cn, lj1994@sjtu.edu.cn, \{jianxiongguo, jiawj\}@bnu.edu.cn}
\thanks{This work is supported in part by the Chinese National Research Fund (NSFC) under Grant 62302048 and 62272050; in part by Guangdong Key Lab of AI and Multi-modal Data Processing, UIC under Grant 2020KSYS007; in part by Zhuhai Science-Tech Innovation Bureau under Grant 2320004002772; and in part by Interdisciplinary Intelligence SuperComputer Center of Beijing Normal University (Zhuhai). (\textit{Corresponding author: Zhiqing Tang.})}
}

\maketitle

\begin{abstract}
Meta computing is a new computing paradigm that aims to efficiently utilize all network computing resources to provide fault-tolerant, personalized services with strong security and privacy guarantees. It also seeks to virtualize the Internet as many meta computers. In meta computing, tasks can be assigned to containers at edge nodes for processing, based on container images with multiple layers. The dynamic and resource-constrained nature of meta computing environments requires an optimal container migration strategy for mobile users to minimize latency. However, the problem of container migration in meta computing has not been thoroughly explored. To address this gap, we present low-latency, layer-aware container migration strategies that consider both proactive and passive migration. Specifically: 1) We formulate the container migration problem in meta computing, taking into account layer dependencies to reduce migration costs and overall task duration by considering four delays. 2) We introduce a reinforcement learning algorithm based on policy gradients to minimize total latency by identifying layer dependencies for action selection, making decisions for both proactive and passive migration. Expert demonstrations are introduced to enhance exploitation. 3) Experiments using real data trajectories show that the algorithm outperforms baseline algorithms, achieving lower total latency.

\end{abstract}


\section{Introduction}
Mobile Edge Computing (MEC) offers substantial compute and storage resources at the network edge to meet Quality of Service (QoS) standards of mobile devices \cite{mao2017survey}. Nevertheless, the lack of trust among edge nodes hinders the consolidation of all computing resources, leading to the emergence of the computing power islands issue \cite{10138337}. Meta computing is suggested as a solution to these issues, presenting a new computing paradigm with zero-trust-based computing modules \cite{10138337, xu2022cloudchain,qi2023latency,li2023online}. By integrating cloud, edge, and other resources and virtualizing networks as meta computers, the management of underlying resources is made transparent to end-users. Services can be deployed through containers on meta computers. When taking into account the user's mobility and the MEC server's limited coverage, the service can be dynamically migrated to a more appropriate MEC server and automatically scaled according to resource demand.

However, challenges arise when making container migration decisions in meta computing. Firstly, there is a need to fully utilize the characteristics of containers to minimize migration cost. Image files must be available locally before running the container \cite{tang2023layer}, containing code, binaries, system tools, configuration files, etc. Otherwise, they need to be downloaded from the registry \cite{registry}. Current methods such as Slacker \cite{slacker}, Cntr \cite{Cntr}, and Pocket \cite{Pocket} aim to reduce startup delays by making significant changes to the application, potentially diminishing the advantages of container isolation. These methods overlook the fact that each image is comprised of layers \cite{layerwork}, which can be shared among multiple images. Researchers have introduced layer matching algorithms for container placement and migration \cite{tang2023layer, lou2023efficient, Tang}. However, the heterogeneity of edge nodes and user mobility in meta computing lead to complex and sparse layer sharing patterns among different tasks. These layer-specific features are both numerous and sparse. Developing a new method for extracting layer features poses a significant challenge that needs to be addressed.

The second challenge is how to make proactive and passive migration decisions jointly. Both proactive migration and passive migration are necessary in meta computing. Proactive migration involves migrating a container to other nodes based on the node resource status \cite{proactive}. Goanon \textit{et al.} \cite{Goanon} propose an architecture that ensures the monitoring system can proactively maintain QoS while managing resource allocation and scaling proactively. Tran \textit{et al.} \cite{Tran} propose a framework that can use application-specific knowledge to determine what data should be proactively migrated to minimize latency and optimize the use of potentially limited resources. Passive migration involves migrating mobile users beyond the communication range, and relocating services based on the user's movement. The concept of companion fog computing is introduced to migrate the fog service to always be close enough to the served device \cite{Carlo}. In meta computing, it is crucial to ensure that the service can only be performed within the topological distance from the edge node. Thus, making migration decisions remains a challenge.

To make migration decisions, several existing works \cite{work4, work5, tang2018migration, tang2024joint,cui2024latency} propose migration solutions using Markov Decision Process (MDP) \cite{MDP} or Lyapunov optimization \cite{Ly} assuming complete system information is known. Meanwhile, Deep Reinforcement Learning (DRL) \cite{DRL} emerges as a promising approach for tackling continuous decision-making challenges. For instance, Lee \textit{et al.} \cite{Lee} introduce a DRL-based algorithm considering CPU, memory, disk, and packet loss utilization to determine migration actions. However, these studies insufficiently account for layer features and fail to address proactive and passive migration decisions.

In this paper, we present a layer-aware Proactive and Passive Container Migration (PPCM) algorithm utilizing proximal policy optimization (PPO) \cite{schulman2017proximal}. To capture layer sharing features effectively, we propose a feature extraction network based on Deep $\&$ Cross Network (DCN) \cite{wang2017deep}. DCN is capable of learning low-dimensional feature crossovers and high-dimensional nonlinear features efficiently compared to other machine learning methods, without the need for manual feature engineering. Additionally, we incorporate expert experience to expedite the training process of the algorithm and enhance the optimization of migration decisions. Experimental results demonstrate the superior performance of our algorithm over baseline algorithms.

The contributions of this paper are summarized as follows:
\begin{enumerate}
    \item We introduce the container migration problem in meta computing for the first time. To efficiently reduce migration and computation costs, we thoroughly incorporate layer sharing information in our system model. We employ a DCN-based network for feature extraction to capture sparse layer sharing features.
    \item For proactive and passive migration decisions, we introduce a PPO-based PPCM algorithm. Leveraging layer sharing features from DCN, this algorithm incorporates expert knowledge to enhance learning speed and decision-making efficiency, ultimately optimizing migration costs in the long run.
    \item Experiments are conducted using real data sets. Based on the results, our algorithm shows significant performance improvements compared to traditional image-based algorithms and layer sharing-based heuristic algorithms.
\end{enumerate}

The paper is structured as follows. Section \ref{sec-related-work} introduces the related work. Section \ref{sec-system-model} and Section \ref{sec-problem} presents the system model and problem formulation. Section \ref{sec-algorithms} provides a description of the PPCM algorithm. Section \ref{sec-evaluation} conducts an experimental evaluation, and Section \ref{sec-conclusion} conclude the paper.

\section{Related Work}
\label{sec-related-work}

\subsection{Service Migration}

When users transition between adjacent or overlapping regions, service migration must consider both the migration cost and the QoS expectations of the migrated users. With a large number of mobile users, diverse edge server types, and the intricate interplay between migration cost and transport, finding an optimal strategy poses challenges. Mirkin \textit{et al.} \cite{Mirkin} introduced a checkpointing and restarting feature for containers in OpenVZ, enabling the inspection and restarting of running containers on the same or different hosts. Docker has streamlined the rapid deployment of Linux applications in recent years \cite{linux}. Machen \textit{et al.} \cite{Machen} presented a layered framework for migrating running applications enclosed in virtual machines or containers, demonstrating the superior performance of containers over virtual machines. Wang \textit{et al.} \cite{Wang} proposed an online approximation algorithm with polynomial time complexity for real-time container-based service migration in a dynamic MEC environment. Montero \textit{et al.} \cite{Montero} advocated for container-based service migration using software defined networking and network function virtualisation to enhance security by relocating security applications from end-user devices to trusted network nodes at the network edge. Notably, all the aforementioned studies focus on service migration with containers but overlook the granular storage of images on layer units, neglecting layer granularity in service migration.

\begin{figure*}[h]
\centerline{\includegraphics[width=7in]{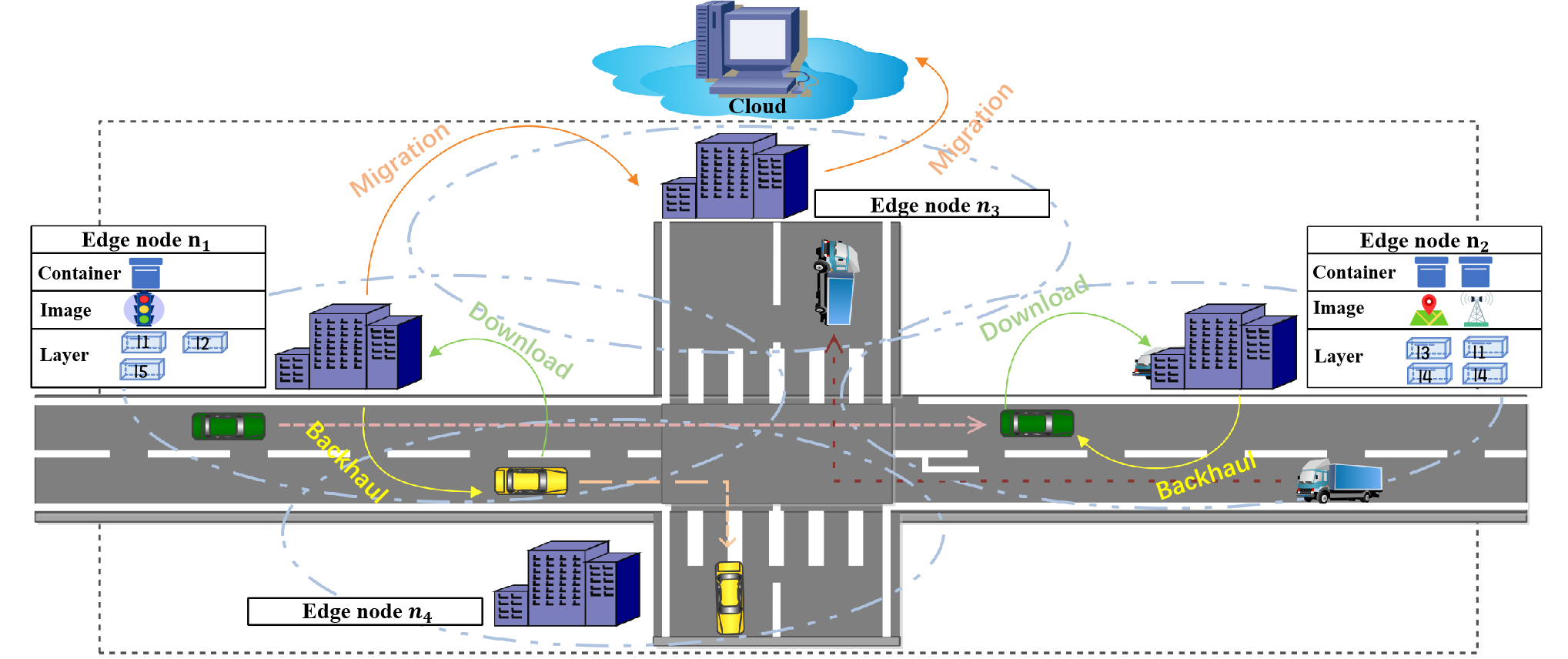}}
\caption{Example of container migration in meta computing.}
\label{fig}
\end{figure*}

\subsection{Layer-aware Scheduling}

Scheduling at layer granularity shows potential. Tang \textit{et al.} \cite{Tang} suggested that utilizing layer-sharing data among containers for multiple users can notably cut migration latency. They introduced an innovative online container migration algorithm to lessen overall task latency. Gu \textit{et al.} \cite{Gu} explored layer-aware microservice placement and request scheduling at the edge, enhancing throughput and hosted microservice volume by sharing layers among nearby images. By sharing layers between these images, the throughput and number of hosted microservices can be significantly boosted. Smet \textit{et al.} \cite{Smet} introduced a heuristic for optimal layer placement, where clients download necessary layers from storage to the server to maximize demand within storage and latency limits. Darrous \textit{et al.} \cite{Darrous} used K-centre optimization for placing container layers in edge nodes, arranging layers by size to minimize maximum retrieval time of container images. Ma \textit{et al.} \cite{Ma} leveraged the hierarchical storage system to reduce filesystem sync overhead and proposed a protocol for real-time migration of edge services for mobile users. Dolati \textit{et al.} \cite{Dolati} presented a polynomial time algorithm based on deterministic and random rounding frames, addressing container layer download paths and locations, VNF chaining during network traffic routing, and orchestrating containerized VNF chains in the edge network. However, none of the above works consider the sharing of information between layers during proactive and passive container migration.

\section{System Model}
\label{sec-system-model}

\subsection{System Model}\label{AA}
Consider a diverse multi-user meta computing system. A group of mobile users are distributed in a specific area $\mathbf{U} = \{u_1, u_2, ... , u_{|\mathbf{U}|}\}$, where $|\mathbf{U}|$ represents the set's size. In this meta computing setup, mobile users can offload their computational tasks to edge node services. Operating on a time-slice model, where a user's location changes only at the start of each time-slice, we can view the issue as a sampling in a continuous time model at various time points $\mathbf{T} = \{t_1, t_2, ... , t_{|\mathbf{T}|}\}$, with a collection of user-generated computational tasks denoted as $\mathbf{K} = \{k_1, k_2, ... , k_{|\mathbf{K}|}\}$. The edge node set is denoted as $\mathbf{N} = \{n_1, n_2, ... , n_{|\mathbf{N}|}\}$, positioned at the edge of the core network. The cloud is treated as an edge node $n_{|\mathbf{N}|+1}$ with limitless computing resources. Each node possesses its CPU frequency $f_n$, bandwidth $b_n$, and storage space $d_n$. For a cloud node, the CPU frequency is $f_{\operatorname{cloud}}$ and the bandwidth is $b_{\operatorname{cloud}}$. To handle the tasks, a variety of containers $\mathbf{C} = \{c_1, c_2, ... , c_{|\mathbf{C}|}\}$ are established and deployed on the edge nodes. Image files are necessary to execute the containers, denoted as $\mathbf{M} = \{m_1, m_2, ... , m_{|\mathbf{M}|}\}$. The images comprise the relevant layers, denoted as $\mathbf{L} = \{l_1, l_2, ... , l_{|\mathbf{L}|}\}$, with each layer $l \in L$ having a size $d_l$.

The set of layers in container $c \in \mathbf{C}$ is denoted as $\mathbf{L_c} = \{x^{l}_{c}|l \in \mathbf{L} \}$, where $x^{l}_{c} = 1$ indicates that container $c$ holds layer $l$, otherwise $x^{l}_{c} = 0$. For each task $k \in \mathbf{K}$ created by the user at time $t$, the CPU resource requested is $p_k$ and the designated container is $c_k$. Following scheduling, the node allocated to this task is represented as $\mathbf{n_k} = \{u^{n}_{k}|n \in \mathbf{N} \cup \{n_{|\mathbf{N}|+1}\}\}$, where $u^{n}_{k} = 1$ implies task $k$ is assigned to node $n$, otherwise $u^{n}_{k} = 0$.

\subsection{Cost}

In a meta computing system, a mobile user $u$ requests service from an edge node at time slice $t$, denoted as $u^{k}_{t}$. Generally, servers are interconnected through stable backhaul links, enabling mobile users to access services via multi-hop communication between edge nodes even when not directly connected to the serving edge node. For optimal outcomes, services should dynamically migrate between nodes.

Migration is split into two parts: proactive migration and passive migration. Proactive migration is triggered when the current node lacks CPU and memory resources, leading to task reclamation and queuing. Resources occupied by the task on the node are then freed. Passive migration occurs when a mobile user surpasses a node's communication range. As each edge node has a limited service range, exceeding it compels the user to switch to another node. If no node meets the criteria during migration, the task is migrated to the cloud. Tasks move dynamically between nodes incurring delays, with the final task time comprising migration delay, computation delay, deployment delay, and backhaul delay.

\textbf{Migration delay}: Migration delay happens when a task moves from one node to another. It has three components. Firstly, the time of movement, i.e., the delay during migration, is defined as:
\begin{equation}
M\left(u, d_t\right)= \begin{cases}0, & \text { if } d_t=0, \\\frac{\operatorname{s}_u^s(t)}{\eta_t^{\mathrm{migr}}}+\sigma_t^{\mathrm{ migr}} d_t, & \text { if } d_t \neq 0,\end{cases}\label{eq1}
\end{equation}
where $d_t$ represents the distance between the current node $u_t^k$ and the previous node $u_{t-1}^k$, $\sigma_t^{\mathrm{m}}$ is a positive coefficient, ${\eta_t^{\mathrm{migr}}}$ denotes the network bandwidth on the migration path, $\operatorname{s}_u^s(t)$ signifies the size of the migrated service. If $d_t=0$, it implies no migration will take place, resulting in a migration delay of 0. Otherwise, if migration occurs, $M\left(u, d_t\right)=0$.

Next, the download duration post migration is a crucial consideration. Each node features a layered download queue. When a new layer requires downloading, it joins this queue and might have to await the completion of another layer. Consequently, the download time for task $k$ equates to the maximum completion time for all the necessary layers. The introduction of $z_{l}^n$ serves to represent the completion time for layer $l$ on node $n$. If the layer is already present or hasn't commenced downloading, then $z_{l}^n = 0$. Otherwise, it is represented as:
\begin{equation}
z_{l}^n = \frac{d_l^u}{b_n}\label{eq2},
\end{equation}
where $d_l^u$ represents the layer size needed by user $u$, $b_n$ stands for the node's bandwidth. In the case of a cloud node, $b_n = b_c$. Hence, the ultimate download time after migration can be formulated as:
\begin{equation}
D(u)=\max _{l \in \mathbf{L}}\left(z_n^l(t) \times x_c^l\right). \label{eq3}
\end{equation}

Third, await execution time. If a task requires migration, it is placed back in the task queue, waiting for selection of a new node, incurring additional wait time. Let the current time slice be $t$ and the time slice entering the task queue be $t_u$, the calculation is:
\begin{equation}
w(u) = t-t_u\label{eq4}.
\end{equation}

Movement and post-migration downloads on each node can happen concurrently, so we only need to consider the maximum of both. The final migration delay can then be calculated as:
\begin{equation}
S = \max(M,D) + \sigma_t^{\mathrm{wait}} w,
\label{eq5}
\end{equation}
where $\sigma_t^{\mathrm{wait}}$ is a positive coefficient.

\textbf{Computation delay}: Mobile users assign computational tasks to nodes for processing, which incurs computation time. At time slice $t$, the task size $\operatorname{s}_u^k(t)$ is offloaded by user $u$, with the task's processing density denoted as $\kappa$. Consequently, the CPU cycles needed to process the offloaded task can be computed as $c_t=\operatorname{s}_u^k(t) \kappa$. Simultaneously, the workload $w_t$ of the node is defined as the multiplication of the total task sizes running at the node and the processing density, i.e., $w_t = \operatorname{s}_u^n(t) \kappa$. The computational power of a node is specified as $f\left(n_t\right)$. For the cloud, $f\left(n_t\right) = f\left(\operatorname{c}_t\right)$. It is assumed that resources are allocated proportionally across nodes, implying that tasks receive computational resources based on their CPU cycle requirements. Therefore, the computational latency of executing the task at time slice $t$ can be determined as:
\begin{equation}
C\left(n_t\right)=\frac{c_t}{\left(\frac{c_t}{w_t+c_t} f\left(n_t\right)\right)}=\frac{w_t+c_t}{f\left(n_t\right)} .\label{eq6}
\end{equation}

\textbf{Deployment delay}: The deployment delay comprises two components. One is the access delay between the mobile user and the node, as defined:
\begin{equation}
A(u) = \frac{\operatorname{s}_u^k(t)}{\rho_t}, \label{eq7}
\end{equation}
where $\operatorname{s}_u^k(t)$ represents the offloading task size ratio for user $u$ at time slice $t$, and $\rho_t$ denotes the wireless uplink transmission rate from the mobile user to the node.

The other aspect is the time taken to deploy to the node for downloading, i.e., the download time after migration $D(u)$. As the mobile user's access and the deployment download can happen concurrently, the deployment delay is the maximum of the two:
\begin{equation}
P = \max(A, D).\label{eq8}
\end{equation}

\textbf{Backhaul delay}: The backhaul delay is the delay between an edge node and the user, defined as:
\begin{equation}
B(u) =\frac{\operatorname{s}_u^k(t)}{\eta_t^{\mathrm{bh}}}+\sigma_t^{\mathrm{bh}} y_t, \label{eq9}
\end{equation}
where $y_t$ represents the distance from the node to the user, $\sigma_t^{\mathrm{bh}}$ is a positive coefficient, and $\eta_t^{\mathrm{bh}}$ denotes the network bandwidth for the return path.

Finally, the overall task duration is defined as:
\begin{equation} 
T_k = S + C + P + B.
\label{eq10}
\end{equation}

\section{Problem Formulation and Analysis}
\label{sec-problem}

\textbf{Constraints}: During migration, decisions must meet various conditions. Firstly, a restriction on the maximum number of containers a node can simultaneously execute is defined as:
\begin{equation}
\left|\mathbf{C}_n(t)\right| \leq C_n, \quad \forall t, \forall n .\label{eq11}
\end{equation}

Moreover, every node is subject to constraints on storage resources and computational capabilities. In the event that, at time slice $t$, the current node's resources fall below a specified threshold, proactive migration becomes necessary. This limitation is defined as:
\begin{equation}
\begin{aligned}
\sum_{l \in \mathbf{L}}\left(1-y_n^l(t)\right) \times d_l &\leq \sigma^{\mathrm{mem}} \times d_n, \quad\quad \forall t, \forall n ,\\
\sum_{l \in \mathbf{L}}{f\left(n_t^l\right)}&\leq \sigma^{\mathrm{cpu}} \times {f\left(n_t\right)}, \quad \forall t, \forall n, \label{eq13}
\end{aligned}
\end{equation}
where $y_n^l(t)$ denotes if layer $l$ is active on node $n$ at time $t$. $\sigma^{\mathrm{mem}}$ and $\sigma^{\mathrm{cpu}}$ are both positive threshold control coefficients.

Third, each task can only be deployed to one node or the cloud, as follows:
\begin{equation}
\sum_{n \in \mathbf{N} \cup\{n|\mathbf{N}|+1\}} u_k^n=1, \quad \forall k .\label{eq14}
\end{equation}

\textbf{Problem formulation}: Our aim is to minimize the total task time of the dynamic meta-computing system while meeting the constraints. Thus, the problem is formulated as:

\newtheorem{problem}{Problem}
\begin{problem} 
$\min T=\sum_{k \in \mathbf{K}} T_k,$
\label{problem}
\begin{equation}
\begin{aligned}
\label{eq:r}
 \text { s.t. } \quad & \text{Eqs.} \  \eqref{eq11}-\eqref{eq14}.
\end{aligned}
\end{equation}
\end{problem}

\begin{figure*}[h]
\centerline{\includegraphics[width=1\textwidth]{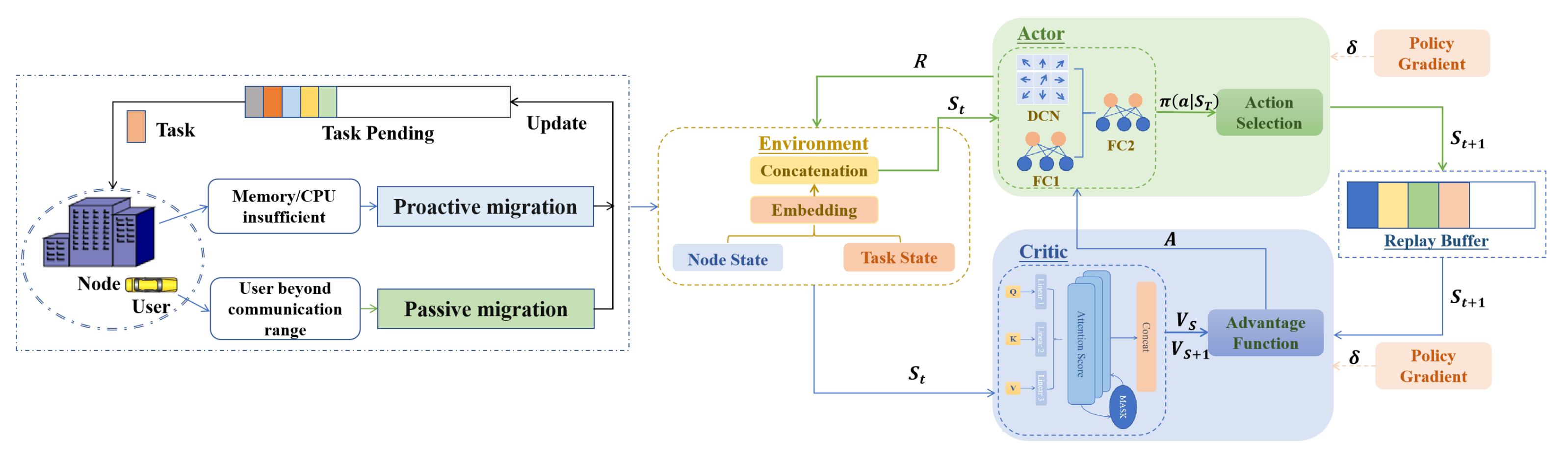}}
\caption{Overview of the layer-aware proactive and passive container migration algorithm in meta computing.}
\label{fig2}
\end{figure*}

Problem \ref{problem} is a complex bin-packing problem that is NP-Hard and can only be tackled heuristically. Traditional algorithms can only iteratively decide based on a set strategy at each time step, lacking insight into the influential factors between adjacent time steps and the ability to adjust strategy as per the meta computing system's dynamics. Making decisions iteratively on a time slice is also notably time-consuming. Problem \ref{problem} meets the MDP criteria by carefully selecting time slices where the first-order transfer probabilities of task resource requirements are quasi-static rather than uniformly spread over an extended period, while also possessing memoryless task and environment updates. In contrast, reinforcement learning algorithms can progressively enhance solutions through continual learning and optimization.

\section{Our Algorithms}
\label{sec-algorithms}

\subsection{Algorithm Settings}

\textbf{State}: The state $S_t$ provides a comprehensive overview of the meta computing, encompassing nodes and tasks.

The node state $s_t^{\mathbf{N}}$ contains the positional coordinates of the edge node $\{L_n^1(t) \cdots L_n^{|\mathbf{N}|}(t)\}$, comprising $\{x_n^1(t) \cdots x_n^{|\mathbf{N}|}(t)\}$ and $\{y_n^1(t) \cdots y_n^{|\mathbf{N}|}(t)\}$. It also includes the distance between the mobile user of the next task and the node $\{o_n^1(t) \cdots o_n^{|\mathbf{N}|}(t)\}$, along with the total size of the next task to be downloaded on the node $\{z_n^1(t) \cdots z_n^{|\mathbf{N}|}(t)\}$. The resource status of the node covers CPU resources $f_n$ and bandwidth $b_n$, as well as the total size of all running tasks on the node $\{\operatorname{s}_u^1(t) \cdots \operatorname{s}_u^{|\mathbf{N}|}(t)\}$. Additionally, the state of each layer owned by the node $\{L_n^1(t) \cdots L_n^{|\mathbf{N}|}(t)\}$ is considered. Therefore, the final node state is represented as:
\begin{equation}
s_t^{\mathbf{N}}=\left\{\begin{array}{ccc}
L_n^1(t) &\cdots & L_n^{|\mathbf{N}|}(t)\\\
o_n^1(t)& \cdots & o_n^{|\mathbf{N}|}(t) \\
z_n^1(t) & \cdots & z_n^{|\mathbf{L}|}(t) \\
\operatorname{s}_u^1(t) & \cdots &\operatorname{s}_u^{|\mathbf{N}|}(t) \\
L_n^1(t) &\cdots &L_n^{|\mathbf{N}|}(t)
\end{array}\right\} \cup \{f_n,b_n\}.
\end{equation}

The task state $s_t^k$ includes the location coordinates $L_u(t)$ of the mobile user, consisting of $x_u(t)$ and $y_u(t)$. It also encompasses the resources required for the task, such as CPU resources $f_u$ and bandwidth $b_u$, along with the total download size $D_u(t)$ needed by the mobile user for the upcoming task. Additionally, it includes the layer state required for the next task $\{L_u^1(t) \cdots L_u^{|\mathbf{L}|}(t)\}$. Therefore, the final task state can be depicted as:
\begin{equation}
s_t^k=\left\{\begin{array}{ccc}
L_u^1(t) &\cdots &L_u^{|\mathbf{L}|}(t)\\
\end{array}\right\} \cup \{L_u(t),f_u,b_u,D_u(t)\}.
\end{equation}
The final representation of the state is:
\begin{equation}
s_t=s_t^{\mathbf{N}} \cup s_t^k .
\end{equation}

\textbf{Action}: The agent performs actions to assign task nodes, so the action space consists of all nodes and the cloud as follows:
\begin{equation}
a_t \in \mathbf{N} \cup\left\{n_{|\mathbf{N}|+1}\right\} .
\end{equation}

\textbf{Reward}: The algorithm aims to maximize reward, while in meta computing, the objective is to minimize total task time, hence the reward is calculated as $r_t=-T_k$. From a long-term perspective, the cumulative reward is:
\begin{equation}
R(\tau)=\sum_{t=0}^T \gamma^t r_t,
\end{equation}
where $ \gamma^t $ is a discount factor ranging from $0$ to $1$.

\begin{algorithm}[h]
    \caption{PPCM}
    \begin{algorithmic}[1]
        \REQUIRE{Initialize policy parameters $\theta$, replay memory $\mathcal{D}_i$, value function parameters $\phi$, expert guidance threshold $\alpha$, task pending $\mathcal{P}$}
        \ENSURE{Selected action $a_t$}
        \FOR{episode $i \leftarrow 0,1,2,...$}
            \STATE Initialize replay memory $\mathcal{D}_i = \emptyset$
            \STATE Initialize task queue $\mathcal{Q} = \emptyset$
            \IF{$i$ \textless  $\alpha_{\text{expert}}$} 
                \FOR{time step $t \leftarrow 0,1,2,...$}
                    \STATE Select action $a_t$ using expert policy
                    \STATE Store transition $(s_t, a_t, r_t, s_{t+1})$ in $\mathcal{D}_i$
                \ENDFOR
            \ENDIF
            \FOR{time step $t \leftarrow 0,1,2,...$}
                \STATE Observe current state $s_t$
                \IF{new task arrives and  $\mathcal{P}$ is not empty }
                    \FOR{each task $k$ in $\mathcal{P}$}
                        \IF{resource insufficient for $k$}
                            \STATE Perform proactive migration for $k$
                        \ELSIF{user beyond communication range for $k$}
                            \STATE Perform passive migration for $k$
                        \ENDIF
                        \STATE Update task status in $\mathcal{P}$
                    \ENDFOR
                \ENDIF
                \STATE Select and execute action $a_t$ using policy $\pi_{\theta}(a_t | s_t)$
                \STATE Observe reward $r_t$ and next state $s_{t+1}$
                \STATE Store transition $(s_t, a_t, r_t, s_{t+1})$ in $\mathcal{D}_i$
            \ENDFOR
            \FOR{training step $j \leftarrow 0,1,2,...$}
                \STATE Estimate advantages using Eq. \eqref{eq-advantage}
                \STATE Compute policy update using Eq. \eqref{eq-policy}
                \STATE Update policy by maximizing Eq. \eqref{eq-objective}
            \ENDFOR
        \ENDFOR
    \end{algorithmic}
    \label{algorithm1}
\end{algorithm}

\subsection{Expert Demonstrations}

In traditional reinforcement learning, the agent seeks to uncover the optimal strategy for maximizing long-term rewards. Typically, the learning process commences with random exploration. However, as the exploration space expands exponentially in meta computing, particularly in scenarios with limited rewards, it can lead to ineffective learning. As suggested by \cite{qiu}, the PPCM algorithm proposes pre-training agents with expert demonstrations gathered through conventional means. This approach involves pre-training agents by first introducing a baseline algorithm based on expert demonstrations, storing the demonstration outcomes in an experience buffer to facilitate agent pre-training, initiating learning through imitation as a warm-start, and subsequently applying traditional reinforcement learning algorithms to optimize the pre-trained strategies. Simultaneously, to prevent excessive influence from the demonstrations on the learned policy, the number of demonstration samples is gradually reduced throughout the pre-training phase \cite{Lee}. The details are outlined in the following section.

\subsection{Migration Policy}

The algorithm overview is depicted in Fig \ref{fig2}. The agent observes node and task states, embeds and connects them, then inputs into the policy network for action selection, migration, and reward generation. The policy network and value function are updated using a policy gradient-based algorithm.

\textbf{Policy network}:
DCN \cite{DCN} is employed in the policy network for extracting inter-dependencies within and across layers. This deep model efficiently learns low-dimensional feature crossovers and high-dimensional nonlinear features while demanding minimal computational resources. Network input features are categorized into dense and sparse features. The sparse feature input comprises the layer state of each edge node, while the dense feature input includes the other states. The cross network layer learns a low-dimensional cross combination of features, depicted as:
\begin{equation}
\mathbf{x}_{l+1}=\mathbf{x}_0 \mathbf{x}_l^T \mathbf{w}_l+\mathbf{b}_l+\mathbf{x}_l.
\end{equation}
where $\mathbf{x}_{l}$ and $\mathbf{x}_{l+1}$ represent column vectors denoting the outputs from the $l$-th and $(l+1)$-th cross layers; $\mathbf{w}_{l}$ and $\mathbf{b}_{l}$ denote the weight and bias parameters of the $l$-th layer. In the deep network section, a fully connected neural network is utilized to learn high-dimensional nonlinear feature cross combinations. The formula is as follows:
\begin{equation}
\mathbf{h}_{l+1}=f\left(W_l \mathbf{h}_l+\mathbf{b}_l\right).
\end{equation}
where $\mathbf{h}_l$ and $\mathbf{h}_{l+1}$ represent the $l$-th and $(l+1)$-th hidden layer; $W_l$ and $\mathrm{b}_l$ are parameters for the $l$-th layer. $f(\cdot)$ denotes the ReLU function. Finally, the outputs of the cross network and deep network are merged to create the combination layer.

\textbf{Value network}: In value networks, we employ a multi-head attention mechanism \cite{multi} to assess state values. This mechanism can merge encoded representation information from various subspaces, analyze the initial input sequences using multiple self-attention sets, and subsequently integrate the outcomes of each self-attention set for a linear transformation to derive the final output results. It is obtained as follows:
\begin{equation}
\begin{aligned}
\operatorname{Attention}(Q, K, V)& =\operatorname{softmax}\left(\frac{Q K^T}{\sqrt{d_k}}\right) V,\\
\operatorname{MultiHead}(Q, K, V) & = \operatorname{Concat}(\operatorname{head}_1, \ldots, \operatorname{head}_{h}) W^O,
\end{aligned}
\end{equation}
where $Q,K,V$ are query matrices, key matrices, and value matrices, respectively. $\operatorname{head} = \operatorname{Attention}(QW_i^Q, KW_i^K, VW_i^V)$.

\textbf{Training}: The PPCM algorithm is designed based on the PPO algorithm. PPO is built upon the policy gradient algorithm to guarantee efficiency with low computational complexity. Given $\pi_\theta$ as a policy with parameter $\theta$, the optimization objective of PPO is as follows:
\begin{equation}
\theta_{k+1}=\arg \max _\theta \mathcal{L}^{P P O}\left(\theta_k, \theta\right). \label{eq-objective}
\end{equation}
The loss function for the policy gradient is defined as:
\begin{equation}
\begin{aligned}
\mathcal{L}^{P P O}\left(\theta_k, \theta\right)= & \underset{s, a \sim \pi_{\theta_k}}{\mathrm{E}}\left[\left(\frac{\pi_\theta(a \mid s)}{\pi_{\theta_k}(a \mid s)} A^{\pi_{\theta_k}}(s, a),\right.\right. \\
& \left.\left.\operatorname{clip}\left(\frac{\pi_\theta(a \mid s)}{\pi_{\theta_k}(a \mid s)}, 1-\epsilon, 1+\epsilon\right) A^{\pi_{\theta_k}}(s, a)\right)\right], \label{eq-policy}
\end{aligned}
\end{equation}
where the function $\operatorname{clip}(x, y, z)=\max (\min (x, z), y)$ restricts $x$ to the interval $[y, z]$, where $\epsilon$ denotes the clip's range. Moreover, PPO employs a generalized dominance estimator for dominance calculation, which can be computed as follows:
\begin{equation}
\begin{aligned}
\hat{A}_t&=\delta_t+(\gamma \lambda) \delta_{t+1}+\cdots+\cdots+(\gamma \lambda)^{T-t+1} \delta_{T-1},\\
\delta_t&=r_t+\gamma V\left(s_{t+1}\right)-V\left(s_t\right),\label{eq-advantage}
\end{aligned}
\end{equation}
where $\lambda$ represents a General Advantage Estimation (GAE) parameter \cite{schulman2015high}, $\delta_t$ signifies the temporal difference error at time step $t$, and $V$ stands for an approximation function.

\begin{figure*}[htbp]
    \centering
    \begin{subfigure}{.45\textwidth}
        \centering
        \includegraphics[width=\linewidth]{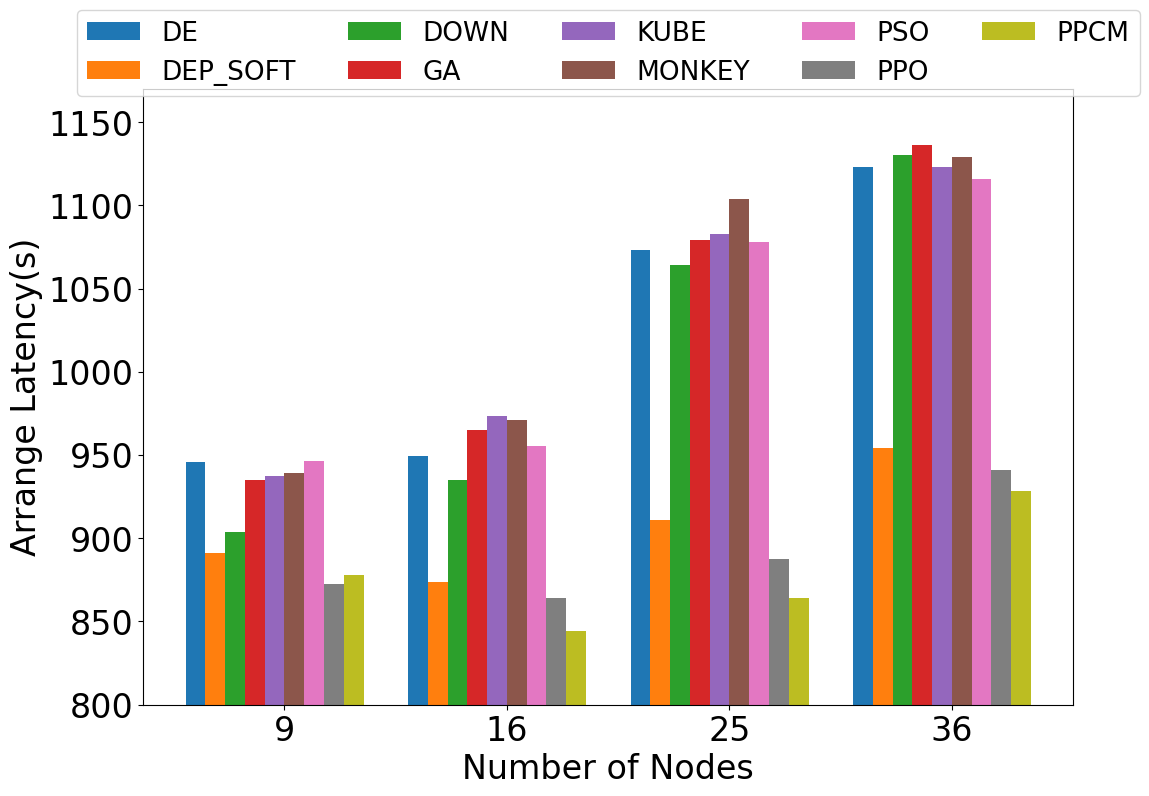}
        \caption{Arrange latency}
    \end{subfigure}%
    \hfill
    \begin{subfigure}{.45\textwidth}
        \centering
        \includegraphics[width=\linewidth]{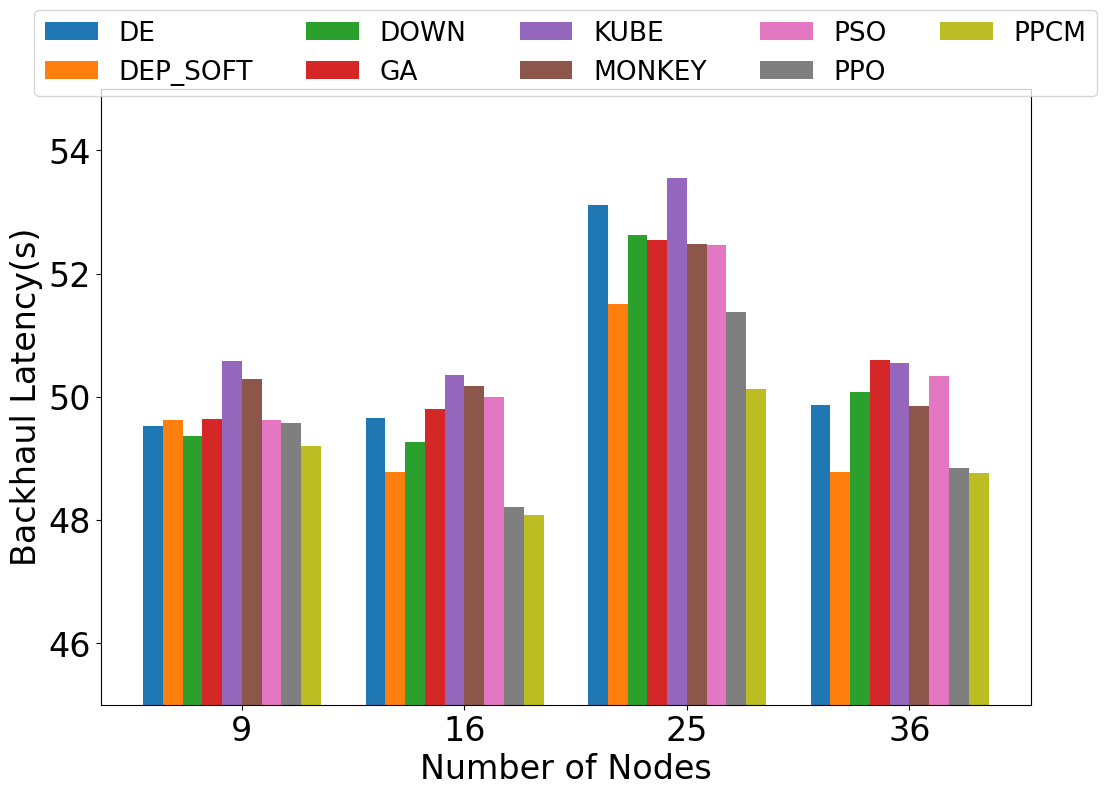}
        \caption{Backhaul latency}
    \end{subfigure}%
    \hfill
    \begin{subfigure}{.45\textwidth}
        \centering
        \includegraphics[width=\linewidth]{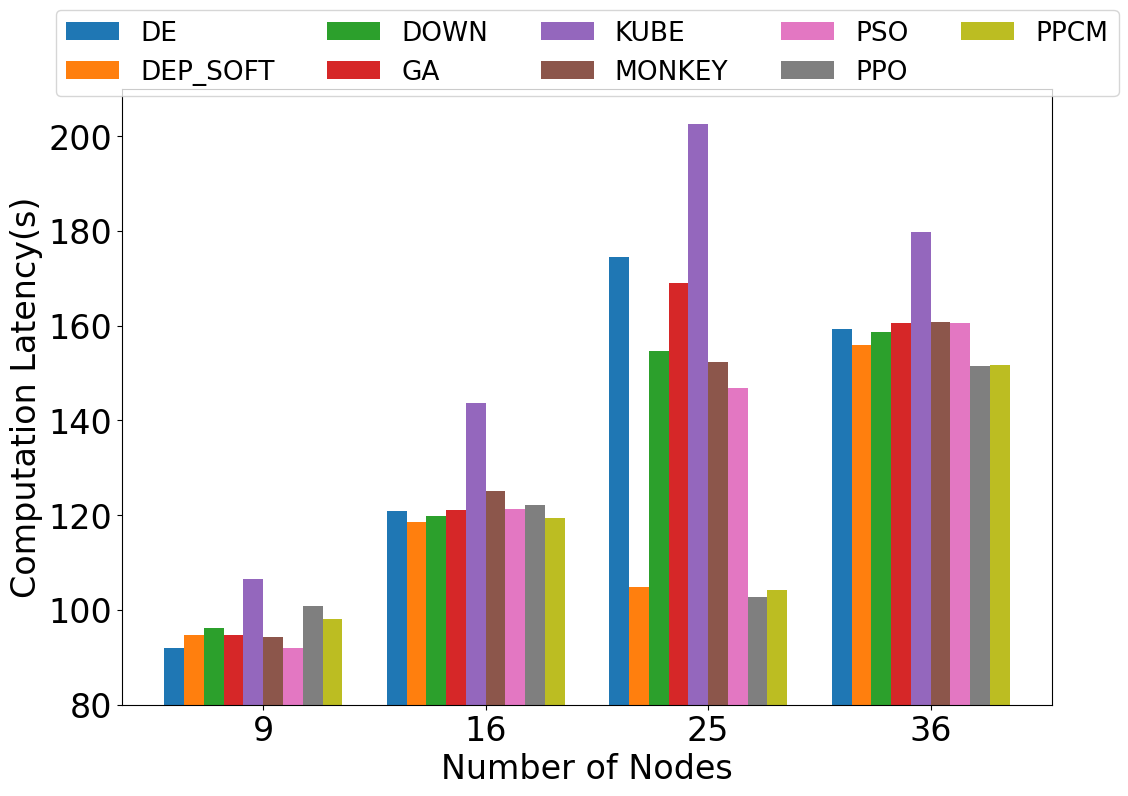}
        \caption{Computation latency}
    \end{subfigure}%
    \hfill
    \begin{subfigure}{.45\textwidth}
        \centering
        \includegraphics[width=\linewidth]{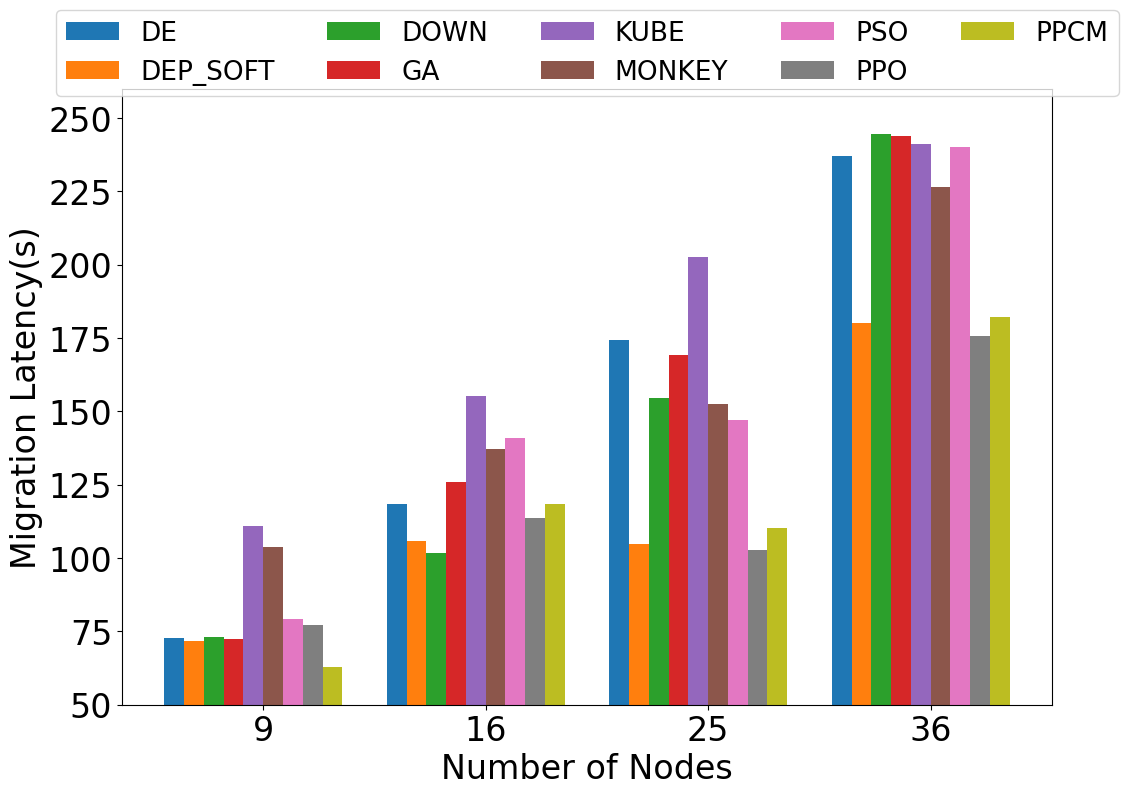}
        \caption{Migration latency}
    \end{subfigure}%
    \caption{Performance with different number of nodes}
    \label{fig5}
\end{figure*}

\begin{figure}
    \centering
    \includegraphics[width=0.9\linewidth]{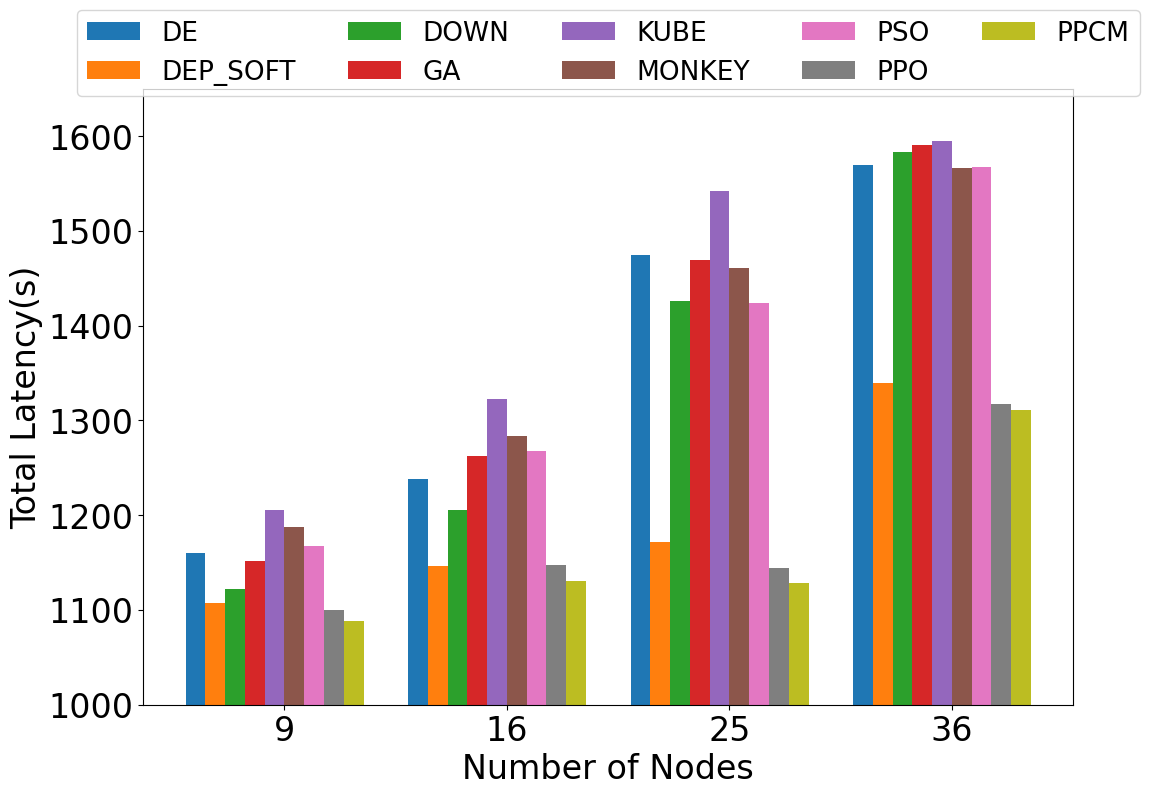}
    \caption{Total latency}
    \label{fig:enter-label4}
\end{figure}

The PPCM algorithm is explained in Algorithm \ref{algorithm1}. Initially, the algorithm initializes the replay memory $\mathcal{D}_i$ for each time point (line 2). Expert demonstrations occur during pre-processing, and at each time step $t$, the expert strategy chooses the action and records the state transition $(s_t, a_t, r_t, s_{t+1})$ in $\mathcal{D}_i$ (lines 4-8). Then, at each time step $t$, the algorithm observes the current state $s_t$ (line 11) and makes a migration type judgment if a new task arrives and the task queue is not empty (line 12). Proactive migration is carried out if the node has low resources (lines 14-15). Passive migration is executed if the user surpasses the communication range (lines 16-17). Following migration, a state update is performed (line 19). The action $a_t$ is chosen based on the current policy $\pi_{\theta}$, executed, and the reward along with the next state are observed, and the state transition is stored in $\mathcal{D}_i$ (lines 22-24). Finally, during the training phase, for each training step $j$ in each episode $i$, the algorithm utilizes the Eq. \eqref{eq-advantage} to compute a policy update (lines 26-29). This update is accomplished by maximizing the objective function using the stochastic gradient ascent algorithm of the Adam algorithm.

\begin{table}[h]
  \centering
  \caption{Hyperparameter Settings of PPCM Algorithm}
    \begin{tabular}{cc|cc}
    \toprule
    Hyperparameter & Value & Hyperparameter & Value \\
    \midrule
    Actor layer type & Dense & Critic layer type & Dense \\
    Learning rate  & 0.0005  & Training epoch & 10 \\
    Batch size & 512 & Optimizer & Adam \\
    Discount $\lambda$ & 0.95 & Discount $\gamma$ & 0.99 \\
    Coefficient $c_h$ & 0.01 & Clipping Value $\epsilon$ & 0.2 \\
    \bottomrule
    \end{tabular}%
  \label{tab:addlabel}%
\end{table}%

\begin{figure*}[htbp]
    \centering
    \begin{subfigure}{.45\textwidth}
        \centering
        \includegraphics[width=\linewidth]{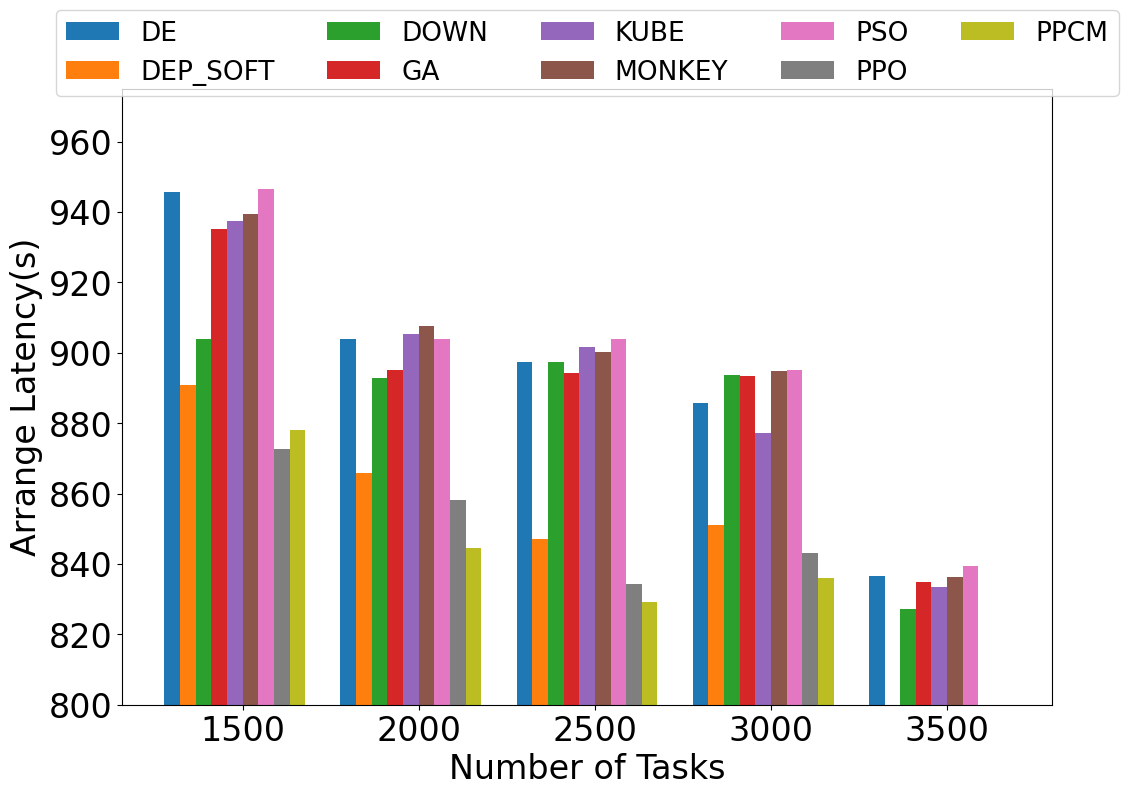}
        \caption{Arrange latency}
    \end{subfigure}%
    \hfill
    \begin{subfigure}{.45\textwidth}
        \centering
        \includegraphics[width=\linewidth]{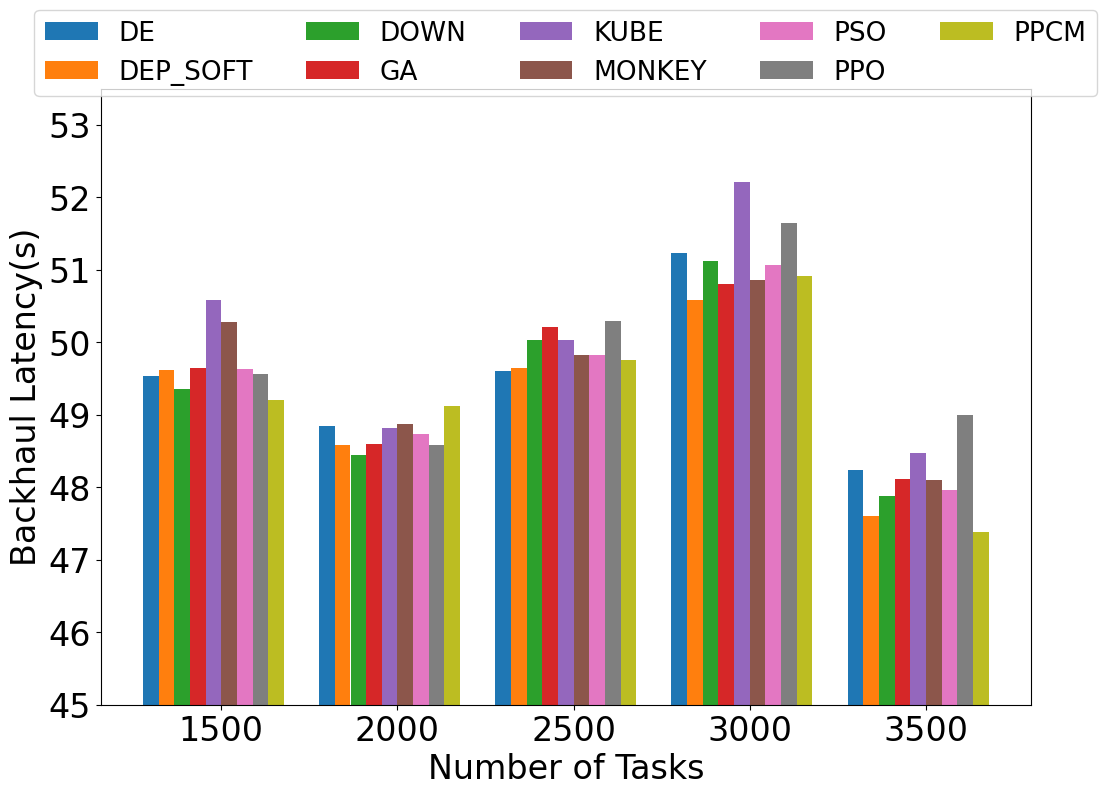}
        \caption{Backhaul latency}
    \end{subfigure}%
    \hfill
    \begin{subfigure}{.45\textwidth}
        \centering
        \includegraphics[width=\linewidth]{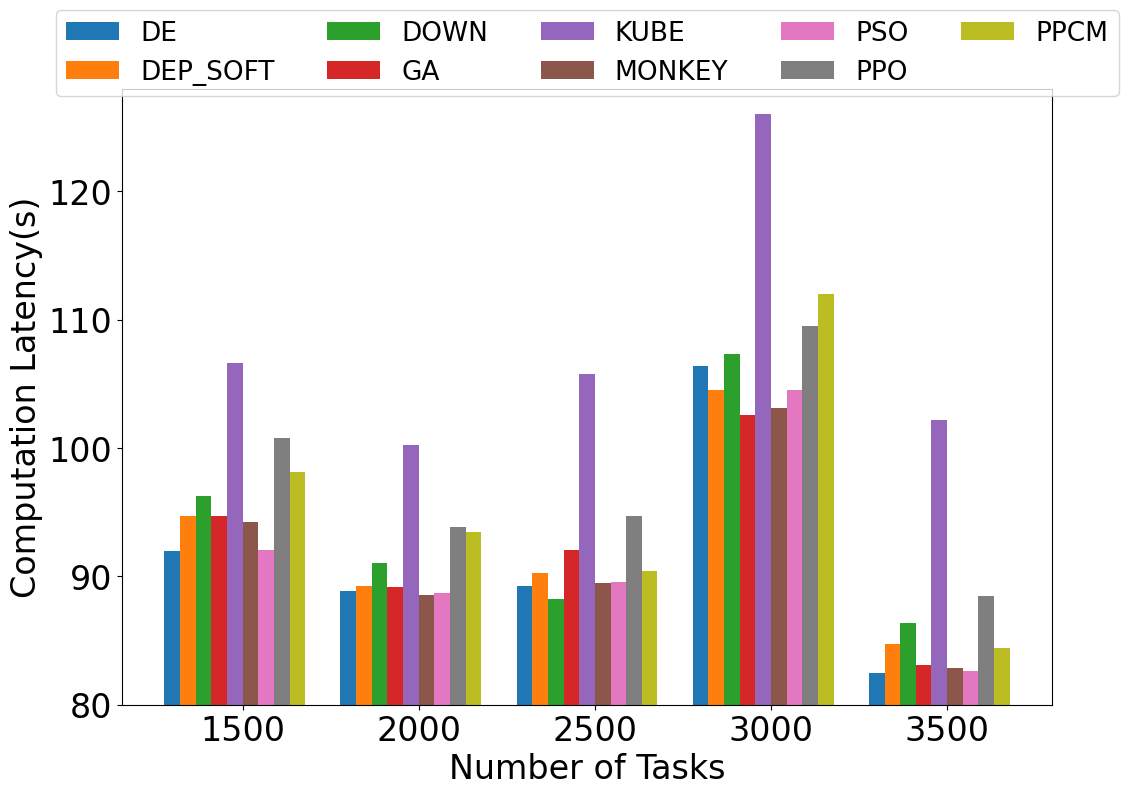}
        \caption{Computation latency}
    \end{subfigure}%
    \hfill
    \begin{subfigure}{.45\textwidth}
        \centering
        \includegraphics[width=\linewidth]{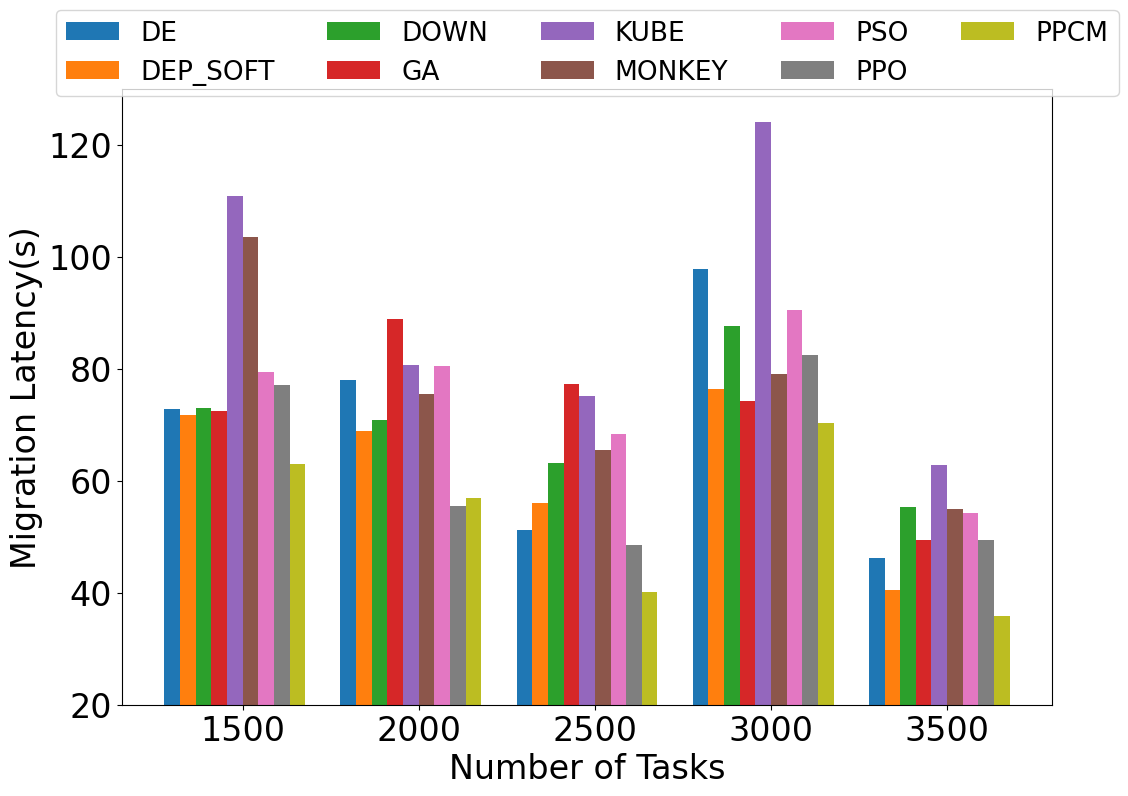}
        \caption{Migration latency}
    \end{subfigure}%
    \caption{Performance with different number of tasks}
    \label{fig7}
\end{figure*}

\begin{figure}
    \centering
    \includegraphics[width=0.9\linewidth]{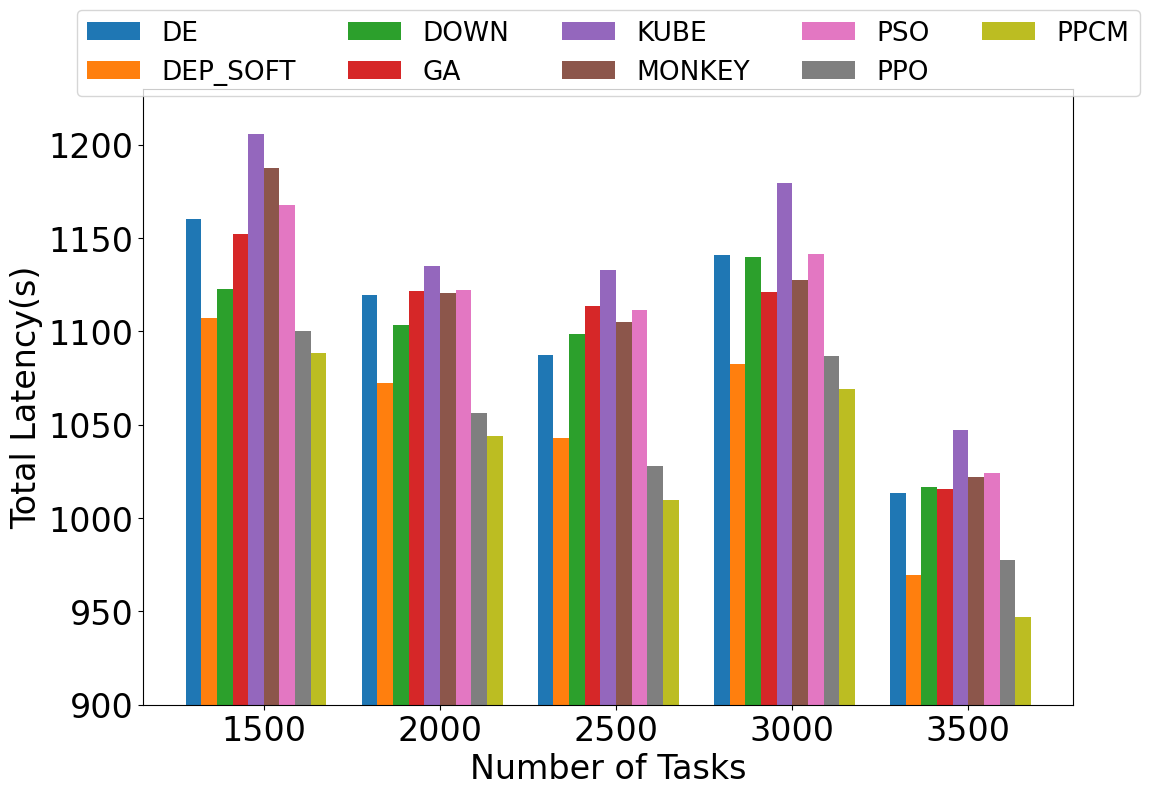}
    \caption{Total latency}
    \label{fig:enter-label6}
\end{figure}

\section{Evaluation}
\label{sec-evaluation}

\subsection{Experimental Settings}

We assess the algorithm using actual mobile trajectories in Rome, Italy \cite{dataset}. We deploy 9 edge nodes in each region, each covering a $\SI{1}{km} \times \SI{1}{km}$ grid with a computational power of $\SI{128}{GHz}$. Typical real-world commercial 5G networks have upload rates of under $\SI{60}{Mbps}$. Hence, in our scenario, the upload rate in each grid is set progressively as $\SI{60}{Mbps}$, $\SI{48}{Mbps}$, $\SI{36}{Mbps}$, $\SI{24}{Mbps}$, and $\SI{12}{Mbps}$. The distance between two edge nodes is calculated using the Manhattan distance. Migration delay varies based on service sizes and network conditions. We assume the service size follows a uniform distribution in the range of $[0.5,100] \SI{}{MB}$, and the migration delay factor follows a uniform distribution in the range of $[1.0,3.0] \SI{}{s/hop}$ during the training phase.

\begin{figure*}[htbp]
    \centering
    \begin{subfigure}{.32\textwidth}
        \centering
        \includegraphics[width=\linewidth]{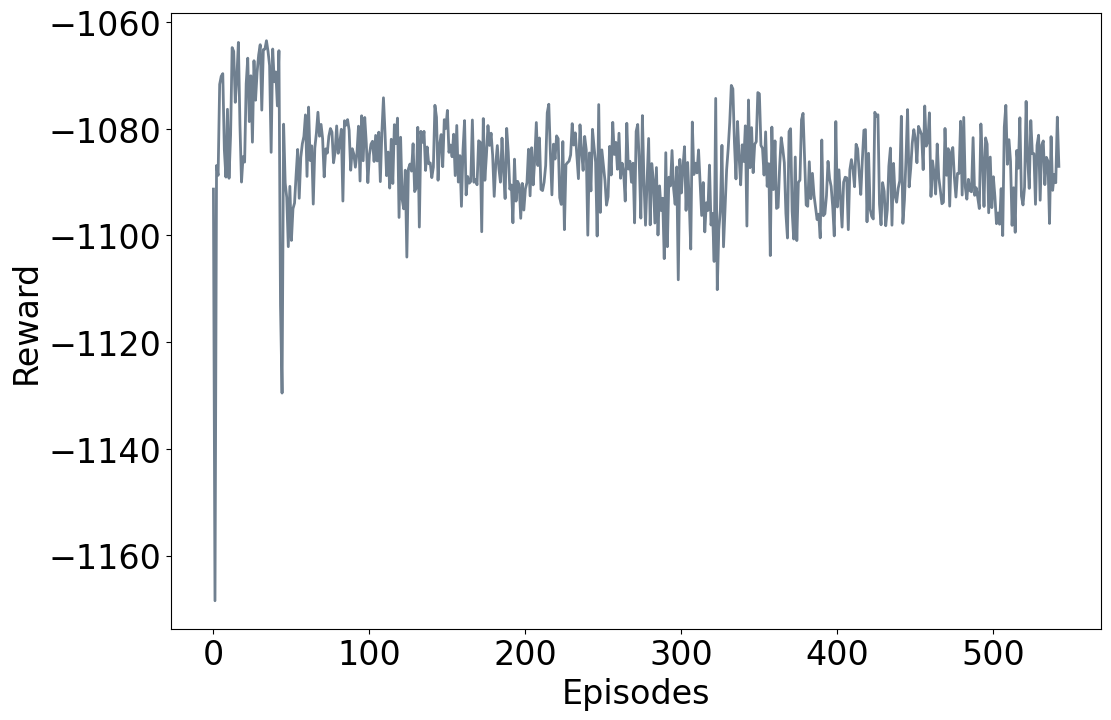}
        \caption{Reward}
        \label{fig:reward}
    \end{subfigure}
    \hfill
    \begin{subfigure}{.32\textwidth}
        \centering
        \includegraphics[width=\linewidth]{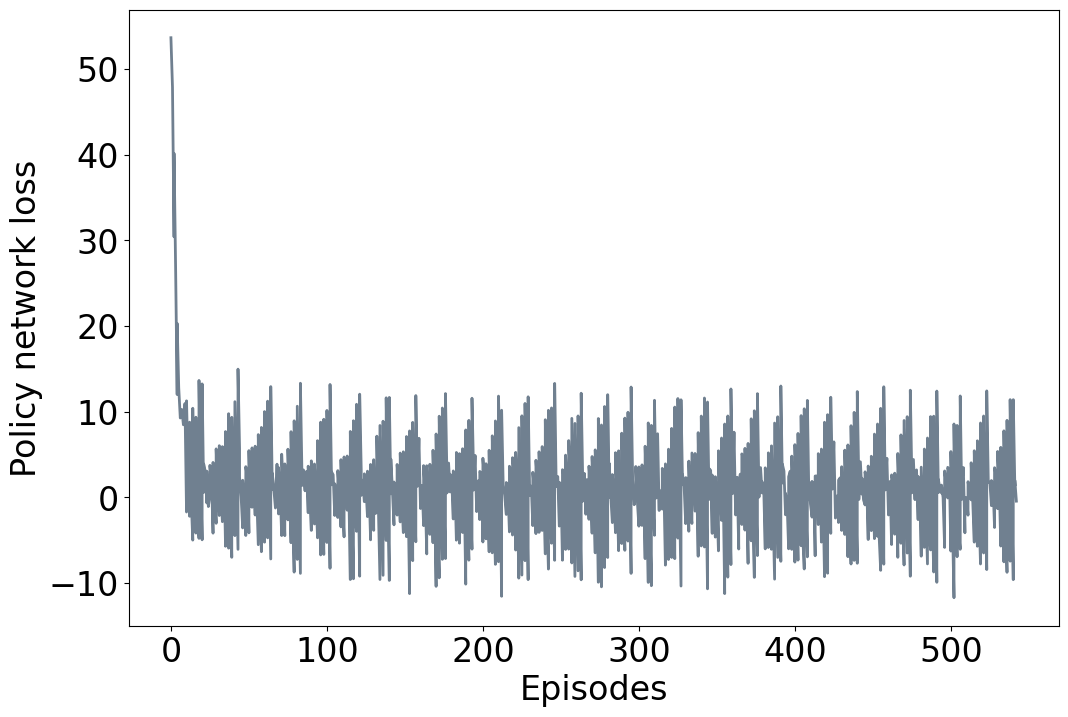}
        \caption{Policy network loss}
        \label{fig:actor}
    \end{subfigure}
    \hfill
    \begin{subfigure}{.32\textwidth}
        \centering
        \includegraphics[width=\linewidth]{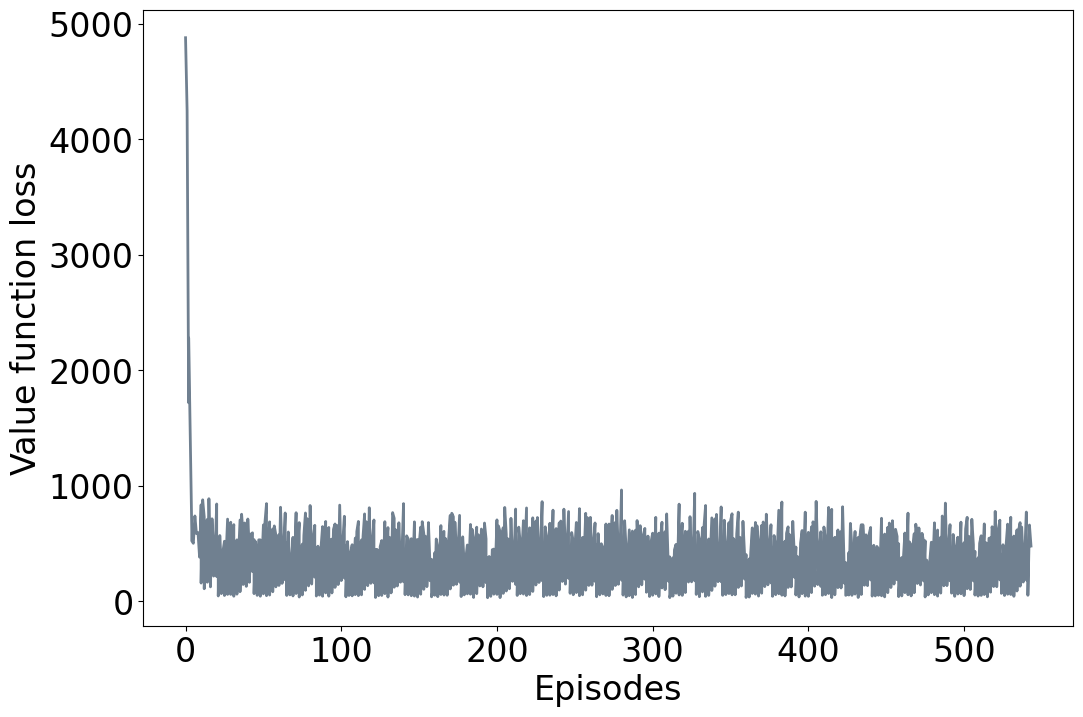}
        \caption{Value function loss}
        \label{fig:critic}
    \end{subfigure}
    \caption{Policy network Loss, value function Loss, and reward of the PPO-Expert algorithm}
    \label{fig6}
\end{figure*}

At each time step, tasks reaching the mobile user and the edge node are randomly drawn from the Poisson distribution. According to empirical data, we evenly allocate the data size for each offloading task within the range of $[0.05,5] \SI{}{MB}$, randomly select the service size from $[0.5,100] \SI{}{MB}$, and uniformly distribute the processing density of the nodes within $[200,10,000]$ cycles. Proactive migration is contingent on resource thresholds, where a positive factor of $0.2$ of the resources acts as the triggering threshold, implying that migration occurs when less than $20\%$ of the resources are available. Node resources are constrained to $[30, 100] \SI{}{GB}$, wired transmission bandwidth ranges from $[800, 1024] \SI{}{Mbps}$, backhaul network bandwidth is within $[400,600] \SI{}{Mbps}$, and backhaul delay factor is fixed at $\SI{0.02}{s/hop}$. The node can support a maximum of $30$ containers. Besides, hyperparameters for the PPCM algorithm are outlined in TABLE \ref{tab:addlabel}.

\textbf{Baselines}: Various baselines are conducted to assess performance. The action space for these algorithms includes both edge nodes and remote clouds. If any nodes lack sufficient resources, the task is offloaded to the cloud. The specifics are outlined below.
\begin{enumerate}
    \item DEP\_SOFT: It is a layer based scheduling algorithm \cite{tang2023layer}. Scores are calculated using local layer size. Simultaneously, a threshold is established, and a node is randomly chosen from nodes surpassing that threshold.
    \item KUBE: This algorithm is image-based and serves as one of the default scheduling algorithms for Kubernetes. It evaluates the distribution of requested images across nodes and assigns a score according to available image sizes, determining scheduling based on this score.
    \item MONKEY: A random algorithm. It picks a single node randomly for scheduling.
    \item DOWN: When calculating the score, the size of the layer is changed to the estimated download time.
    \item GA: It is a meta-heuristic algorithm inspired by the process of natural selection that relies on biologically inspired operations \cite{tang2023layer}.
    \item DE: Differential Evolution optimizes a problem by iteratively trying to improve candidate solutions with respect to a given quality metric \cite{tang2023layer}.
    \item PSO: Particle Swarm Optimisation is a meta-heuristic algorithm. It solves a problem where there is a swarm of candidate solutions and moves them in the search space according to a simple mathematical formula \cite{tang2023layer}.
    \item PPO: PPCM algorithm without expert demonstrations.
\end{enumerate}

\subsection{Experimental Results}

\textbf{Performance with different number of nodes}. Figs. \ref{fig5} and \ref{fig:enter-label4} illustrate the total delay as the number of nodes scales. The data indicates a noticeable trend: as the node number rises, the average task delay also increases. This phenomenon stems from the escalating cumulative delay caused by migrations. Consequently, the overall task delay grows with the expanding node number. In terms of total delay relationships, the hierarchy stands as follows: PPCM \textless PPO \textless DEP\_SOFT \textless DOWN \textless PSO \textless DE \textless GA \textless MONKEY \textless KUBE, with potential minor deviations in the intermediary algorithms. Notably, PPCM consistently outperforms other algorithms irrespective of node number. Specifically, it achieves a reduction in total delay compared to all baseline algorithms by an average of 18\%.

\textbf{Performance with different number of tasks}. As the number of tasks increases, task delay variation is depicted in Figs. \ref{fig7} and \ref{fig:enter-label6}. Results indicate that the PPCM algorithm exhibits superior performance. Among the baseline algorithms, DEP\_SOFT consistently delivers the best results, while the KUBE algorithm shows the worst performance. Overall, with an increasing number of tasks, the performance hierarchy based on total delay is PPCM \textless PPO \textless DEP\_SOFT \textless DOWN \textless GA \textless DE \textless MONKEY \textless PSO  \textless KUBE. Despite minor differences among the middle baseline algorithms, the PPCM algorithm reduces the total delay by an average of 7\% compared to the baseline algorithms.

\textbf{Performance of PPCM algorithm}. Fig. \ref{fig6} illustrates the convergence of the PPCM algorithm. The policy network loss and value function loss initially have high values during training. Nevertheless, with the progression of training steps, both decrease swiftly and stabilize around a specific value, signifying the algorithm's convergence.

\section{Conclusion}
\label{sec-conclusion}

In this paper, we have modeled the layer-aware proactive and passive container migration problem in meta-computing. To efficiently reduce migration and computation costs, we thoroughly incorporate layer sharing information. For proactive and passive migration decisions, we introduce a PPO-based PPCM algorithm. We employ a DCN-based network for feature extraction to capture sparse layer sharing features. Expert knowledge is incorporated to enhance learning speed and decision-making efficiency, ultimately optimizing migration costs in the long run. Experimental results have shown the effectiveness of our proposed PPCM algorithm. Future work will consider the auto-scaling problem of containers in meta computing.

\bibliographystyle{IEEEtran}
\bibliography{ref}
\end{document}